\documentstyle{article}%
\begin{document}

\title{Looking for packing units of the protein structure}
\author{Wei-Mou Zheng${^{*,1}}$, Hui Zeng$^1$, Dong-Bo Bu$^2$, Ming-Fu Shao$^2$,
Ke-Song Liu$^1$, Chao Wang$^2$\\
\small ${^1}${\it Institute of Theoretical Physics, Academia
Sinica, Beijing 100190, China}\\
\small ${^2}${\it Institute of Computing Technology, Academia
Sinica, Beijing 100190, China}\\
\small ${^*}$ To whom correspondence should be addressed. }
\date{}
\maketitle

\begin{abstract}
Lattice-model simulations and experiments of some small proteins suggest that folding is
essentially controlled by a few conserved contacts. Residues of these conserved contacts
form the minimum set of native contacts needed to ensure foldability. Keeping such
conserved specific contacts in mind, we examine contacts made by two secondary structure
elements of different helices or sheets and look for possible `packing units' of the
protein structure. Two short backbone fragments of width five centred at the $C_\alpha$
atoms in contact is called an H-form, which serves as a candidate for the packing units.
The structural alignment of protein family members or even across families indicates that
there are conservative H-forms which are similar both in their sequences and local geometry,
and consistent with the structural alignment. Carrying strong sequence signals, such packing
units would provide 3D constraints as a complement of the potential functions for the 
structure prediction.

\end{abstract}

\noindent {\bf Key words:} Protein structures; Fragment packing;
Packing units.
%\leftline{PACS number(s): 87.10.+e,02.50.-r}
\bigskip
\section{Introduction}

Assessing structural similarity and defining common patterns through
protein structure comparison is important in functional
and evolutionary studies of proteins.
Commonly occurring structural motifs provide insight into the conservation
of protein structure, the types of structural interactions preferred in
nature, and the relationships among sequence, structure, and function.
Structural motifs are more sensitive in finding distantly related homologs
than structural alignment methods.\cite{brutlag04}
Local structural motifs consist of a
relatively small number of successive residues.
Many sequence segments often adopt a single or just a few local structures
although interactions associated with a long sequence separation may
affect these local structures.\cite{sander06}
The local structures adopted by the most closely related short
sequence segments can be extracted from protein structure databases;
sequence information is useful for prediction of local structural motifs.
Local structure predictions have been successfully
incorporated into fold recognition or fold prediction methods.\cite{baker01}
As a direct extension of short local structures, supersecondary structural
elements have been exhaustively classified, which turn out to be
sufficient to describe all known folds, either common or novel.\cite{smotif10}

Commonly occurring global motifs usually consist of several secondary
structure elements (SSEs) that are not specific to any single fold or
protein family. Thus, they are capable of representing
substructures of most protein structures although
these proteins may have little in common as a whole, both in terms of
structure and sequence.
Detection of global structural similarity hiding in globally dissimilar
structures is complicated by the presence of strong local structural
similarities.
The fold definition concerns only the architecture and topology of major
SSEs without consideration of subtle
differences in 3D coordinates.\cite{tops}
Some methods for global motifs take into account, besides the match of SSE
types and topology, also the handednesses of connections between
SSEs, coordinates of SSE starts and ends, types of interactions between
SSEs, $\beta$-sheet definitions and other features.\cite{prosmos}

The task of structure classification is quite different from that of
structure prediction.
An {\it ab initio} method for protein structure prediction assumes that
all the information of a structure is contained in its sequence of amino
acids, and the native structure has the lowest free energy, but to locate
the native state among native-like structures is exceedingly difficult.
Although Rosetta's potential function contains `sequence independent terms'
of secondary structure vector interaction there is still a lack of good
distance constraints based on the sequence information to fix the spatial
arrangement among SSEs.\cite{baker99}

It is observed that
many protein structures may share a local substructure which consists
of several short backbone fragments closely surrounding a particular
amino acid as the center. A library of so-called `local descriptors
(LDs)', which is general enough to allow assembly of protein structures,
has been constructed.\cite{ld09}
Such local protein structure representations
incorporate contacts between residues (of long-range in sequence separation)
to characterize local neighborhoods of amino acids including short and
long-range interactions. Furthermore, it is possible to identify meaningful
sequence similarity in groups of such LDs, and then hopefully to describe
the sequence-structure relationship within each group in the folding space.

Lattice-model simulations and experiments suggest that folding is essentially
controlled by a few conserved contacts, which form very early in the folding
process, and, after their assembling together, lead to the folding core of the protein.
Residues of these conserved contacts form the minimum set of native contacts
needed to ensure foldability. Their mutations have quite dramatic effects on the
stability of the transition state nucleus and the folding kinetics. While these
conserved residues form most of their contacts in the transition state,
others only do so on reaching the native conformation.\cite{nature,broglia}
In other words, the contacts most responsible for foldability and stability
are most conservative,
and hence carry the strongest sequence information. On the contrary, the others
are `dragged' by the former into the finally compact shape.

Keeping such conserved specific contacts in mind, we look for what we call
`packing units' of the protein structure. As the folding core is our concern,
we focus on SSEs. Each unit will contain only two SSEs, being minimal. We require
that there is at least a `contact' between the two SSEs. That is, a pair of
residues, each on one of the two separated SSEs, meet some distance criterion,
say with two C$_\alpha$ atoms being within $8.5\AA$. Two short backbone
fragments (of width 5) centred at these C$_\alpha$ atoms in contact, which
will be called an H-form, serve as a candidate for the packing units. For
strands on the same $\beta$-sheet, there are many nearby `contacts' of
hydrogen binding. They are less responsible for the overall packing than those
coming from different sheets or between sheets and helices.
We shall ignore the contacts within a sheet when considering H-forms.
For an H-form to be a packing unit, it should appear in certain commonly occurring
global motif to satisfy geometric specificity, moreover its sequence is highly
conservative. We shall discuss the characterization of packing units, their
identification from structure databases, and their relation to other
substructure motifs.

\section{Methods: From H-forms to packing units}

The two SSEs of an H-form are from different helices or sheets. %Denote the two
%fragment sequences by $a_{-2}a_{-1}a_0a_1a_2$ and $b_{-2}b_{-1}b_0b_1b_2$ (with
%the former close to the N terminus).
The N- and C-terminal SSEs of an H-form are referred as SS1 ($a_{-2}a_{-1}a_0a_1a_2$)
and SS2 ($b_{-2}b_{-1}b_0b_1b_2$), respectively. We require that the DSSP states of
$a_0$ and $b_0$ belong to $\{B, E; H, I, G\}$.\cite{dssp} We make the reduction:
$\{B, E\}\to E$ and $\{H, I, G\}\to H$. Some $a_i$ and $b_j$ other than $a_0$
and $b_0$ are allowed to be on loops. We require further that $a_0$ and $b_0$
belong to different helix or sheet according to DSSP annotation.

\subsection{Characterization of the local geometry for an H-form}

An H-form is a geometric object in 3D space. We first need to
determine the axis direction of an SSE. For a long SSE, its axis direction can
be defined as the line that has the minimal sum of distances to all the
$C_\alpha$ atoms of the SSE. This does not work for our short fragments here.
We determine the axis direction by fitting the given four $C_\alpha$ atoms
$a_{-2}a_{-1}a_0a_1$ (or $b_{-2}b_{-1}b_0b_1$) to the `standard helix' which
is described by
\begin{equation}
x=a\cos\theta ,\quad y=a\sin\theta ,\quad z=b\theta ,
\end{equation}
where $a$ is the radius, and $2\pi b$ the pitch. A strand can be viewed as an
extended helix. Denote the position vectors of four $C_\alpha$ atoms
$a_{-2}a_{-1}a_0a_1$ by ${\bf r}_1$, ${\bf r}_2$, ${\bf r}_3$ and ${\bf r}_4$,
respectively. Let ${\bf u}$ be the unit vector of the axis, if fragment
$a_{-2}a_{-1}a_0a_1$ forms a helix. Thus, we have
\begin{equation}
({\bf r}_2-{\bf r}_1)\cdot {\bf u}\equiv {\bf r}_{21}\cdot{\bf u}=b, \quad
({\bf r}_{32}-{\bf r}_{21})\cdot {\bf u}=0,
\end{equation}
This means that vector ${\bf r}_{32}-{\bf r}_{21} = {\bf r}_3-2{\bf r}_2+
{\bf r}_1$ is perpendicular to ${\bf u}$. For a general case of four
successive points not in a plane, ${\bf u}$ is then determined by
\begin{equation}
{\bf u} \propto ({\bf r}_4-2{\bf r}_3+{\bf r}_2)\times
({\bf r}_3-2{\bf r}_2+{\bf r}_1).
\end{equation}
For the in-plane case, we may take ${\bf u}$ along
$({\bf r}_4-2{\bf r}_3+{\bf r}_2)- ({\bf r}_3-2{\bf r}_2+{\bf r}_1)$. Of course,
such a vector ${\bf u}$ has no direct meaning for a fragment on a loop, which is
not in our concern here.

Let ${\bf r}_{ba}$ be the vector from $C_\alpha$ atom of $a_0$ to that of $b_0$,
and ${\bf u}_a$ and ${\bf u}_b$ be axes of fragments $a_{-2}a_{-1}a_0a_1$ and
$b_{-2}b_{-1}b_0b_1$, respectively. The relative orientation of the H-form is
described by the angles
\begin{equation}
\theta_a =\arccos \left[\frac{{\bf u}_a{\bf\cdot}{\bf r}_{ba}}{|{\bf r}_{ba}|}\right],\
\tau_{ab} = {\rm sgn}[{\bf u}_a{\bf\cdot}({\bf r}_{ba}\times {\bf u}_b)]\cdot
\arccos \left[\frac{({\bf u}_a\times {\bf r}_{ba}){\bf\cdot}({\bf r}_{ba}\times {\bf u}_b)}
{|{\bf u}_a\times {\bf r}_{ba}|\cdot |{\bf r}_{ba}\times {\bf u}_b|}\right],\
\theta_b =\arccos \left[\frac{{\bf u}_b{\bf\cdot}{\bf r}_{ba}}{|{\bf r}_{ba}|}\right].
\end{equation}
Thus, the local geometry of the H-form is characterized by $(d\equiv |{\bf r}_{ba}|;
\theta_a,\tau_{ab}, \theta_b)$. These quantities are intrinsic, and independent of
the reference frame. Another important feature of an H-form is its
sequence separation $\ell$ which may be taken as the difference between the
site indices of $a_0$ and $b_0$. Sometimes other features like solvent
accessibility can be also considered.

Generally, side-chains are more responsible for contact packing. It is often to
consider $C_\beta$ atoms or some representative points for distance criteria.
However, their information is not directly available in early stages of structure
prediction. The contact statistics are qualitatively, even quantitatively, similar
when using either $C_\alpha$ or $C_\beta$ atoms.\cite{contact} Thus, we consider
only $C_\alpha$ atoms to facilitate treatment and to make the obtained results easily
and widely applicable.

\subsection{Structure alignment based on a pair of similar H-forms}
An H-form is only an object of local geometry. A packing unit has to be a commonly
occurring motif which exhibits structural and sequential similarities with other
members in the same motif, and at the same time occurs as a part of certain structure
alignments. That is, a packing unit tends to be aligned together with some other SSEs
in two structures.
When inspecting a group of closely related structures, {\it e.g.} a SCOP family,
for packing units, their good candidates can be found simply by structure alignment.
When a pair of similar H-forms are from two distant structures, an ordinary
structure alignment tool usually does not work. For this purpose, we design a tool
for structure alignment based on a pair of similar H-forms. This tool uses the
`zoom-in' technique of our ClePAPS.\cite{clepaps} We determine the transform based
on the two H-forms to superimpose the two structures they belong to. At a large
threshold for coordinate deviation, we enlarge the list of correspondence, which
originally consists of only the H-form pair. (In doing this, an efficient way is to
use our conformational letters.) We then update the transform based on
the enlarged correspondence, and use a more stringent criterion for deviation cutoff
to again update the correspondence. This is an iteration. Usually, three iterations
are enough for judging whether the H-form pair occurs as a part in a structure
alignment containing also H-forms other than the given pair.

Of course, the final alignment depends on the deviation cutoffs for `zoom-in'.
Generally, the `zoom-in' helps to escape from a local trapping, and ensures the
alignment found is more or less grobal. In the extreme case when the pair of H-forms
for the initial alignment are supported by no other SSEs, the initial alignment would
survive iterations only at rather stringent deviation cutoffs.

\section{Results and discussion}
\subsection{CI2 as a case study}

%d2ci2i_     class: Alpha and beta proteins (a+b)
%fold/superfamily/family: CI-2 family of serine protease inhibitors
Chymotrypsin inhibitor 2 (CI2, PDB-ID 2ci2) was taken as the target
conformation for a detailed examination in Ref.~\cite{nature}.
Two residues were considered to be in contact if the distance between their $C_\beta$
atoms ($C_\alpha$ for G) was $\leq 7.5\AA$ and if they were more than two units
apart from each other along the sequence. It was inferred that A16, I20, L49, V51
and I57 to belong to the conserved folding nucleus. (The numbering of residues here
is shifted by 19 from theirs, {\it e.\ g.} A16 here would be their A35.) The
protein belongs to Pfam protein family PF00280, whose seed alignment contains four
known structures.\cite{pfam}

Since CI2 is a small molecule of Length 65, a loose distance threshold $8.5\AA$ is
taken for H-forms. We find seven H-forms in CI2 as listed in Table 1. Residue I57 is
on a loop, so does not appear. Besides the remaining four inferred sites, there are several
extra sites, {\it e.\ g.} L8 in the first H-form. This contact formed by L8 and A16
is responsible for the packing of two helices, which, connecting through a short turn
(of width 4), compose a supersecondary structure. We shall discuss other sites later.

A known
structure in the seed alignment of PF00280 is one with PDB-ID 1vbw. It has 21 H-forms.
The DALI pairwise structure alignment \cite{dali} between 2ci2 and 1vbw is mainly a shift
of 3, {\it e.\ g.} A16 of 2ci2 aligns against A19 of 1vbw. All the H-forms of 1vbw which
have their corresponding H-forms of 2ci2 in the alignment are listed in Table 2, where
listed are also the BLOSUM62 similarity scores of the aligned fragments. The fragment
of V13 of 2ci2, which is the SS1 of the second, third and fourth
H-forms in Table 1, has a rather low similarity score ($-4$) with its aligned partner
in 1vbw, so the fragment of V13 is not very conservative and V13 turns being not among
the inferred sites. %Residue V13 contacts with V51 in the fourth H-form, which has a
%rather large difference in $\tau$ from its counterpart in 1vbw.
The similarity scores between
2ci2 and 1vbw for the sixth H-form are both high, but the difference in distance $d$
is large, indicating a distortion between the two H-forms. Finally, only the fifth and
seventh H-forms exhibit both sequential and structural similarities between the two
proteins. These two H-forms involve sites A16, I20, L49 and V47. For V47 to be a `hot'
site is supported by both the sequence design entropy and alignment entropy given in
Ref.~\cite{nature}. We have further examined another known structure 1mit of the PE00280
seed alignment. It has 28 H-forms. The DALI pairwise structure alignment between 2ci2 and
1mit is mainly a shift of 4. All the H-forms of 1mit which have their corresponding
H-forms of 2ci2 in the alignment are listed in Table 3. Since 1mit has only one helix
no counterpart of the first H-form of 2ci2 exists in 1mit. As for the other H-forms,
the situation is very similar to that of 1vbw. It should be mentioned that the third
H-form shows both sequential and structural similarities between 1vbw and 1mit, so there
is still a possibility for the H-form to be identified as a packing unit after more
structures are inspected.

There are pairs of H-forms between 2ci2 and 1vbw or 1mit which show both sequential
and structural similarities, but conflict with the structure alignment. Two examples are\\
{\tt \phantom{1}2ci2 K17:I29 EAKKV - AQIIV $\sim$ 1vbw K20:V52 AAKAV - VRVWV \quad
(11 5; -0.2, -0.5 0.0  0.4)}\\
{\tt \phantom{22ci2 K17:I29 EAKKV - AQIIV }$\sim$ 1mit K21:I53 VAKAI - VRIWV
\quad (\phantom{1}9 6; -0.1, -0.4 0.5  0.6)}\\
where at the end of each line the similarity comparison is given in the format (two sequence
similarity scores; distance difference, and differences in three angle
($\theta_a,\tau_{ab},\theta_b$)). Such pairs of H-forms usually have quite diverse sequence
separation. The possible physical rationale for occurrence of such correspondence might
be the adjustment of SSEs to make an optimal physical interaction in a later packing stage.
Generally, if an H-form of 2ci2 is compared with H-forms of 1vbw or 1mit, there is
no similarity either in subsequences or in local substructure geometry. A few of pairs
of H-forms might have one similarity, but seldom exhibit both. There is a good chance for
those similar in both sequence and structure to be consistent with the structure
alignment.

\subsection{Inspecting a SCOP super-family for packing units}

Let us inspect six domains from SCOP-40 super-family d.122.1, which contains three
families with at least two members. Two domains
d1y8oa2 and d1gkza2 belong to family d.122.1.4. The CATH domain, which consists of
d1y8oa2 and C-terminus of 1y8o, is about a hundred longer than d1y8oa2. The alignment
between 1y8o and 1gkz shows that the C-terminus contributes to the alignment. We shall
refer d1y8oa2 to the longer CATH domain for our analysis here. Domains d1y8oa2 and
d1gkza2 have 124 and 74 H-forms, respectively. Taking the similarity criteria: 1)
BLOSUM62 sequence similarity scores $\geq 0$ for both fragments; 2) $\Delta d\leq
1.5\AA$, $\Delta\theta\leq 0.6$, and $\Delta\tau\leq 0.8$, we find 80 pairs of
`similar' H-forms, among that 54 pairs coincide with the structure alignment.
(Without requiring the sequence similarity there would be 698 pairs of H-forms
similar in geometry.)
Many of these pairs are clustered in their positions. For example,

\smallskip\quad\parbox{15.5cm}{\tt
d1y8oa2: 31 MLCEQ 73 LFKNS; 31 MLCEQ 76 NSMRA; 31 MLCEQ 77 SMRAT;\\
d1gkza2: 26 RLCEH 62 LLKNA; 26 RLCEH 65 NAMRA; 26 RLCEH 66 AMRAT.}

\smallskip\noindent Here the format for an H-form is: `center position of
fragment 1', `sequence of fragment 1', `center position of fragment 2',
`sequence of fragment 2'. Since the two domains are
very close in structure and sequence, we cannot simply tell which H-forms are
leading, and which are dragged-in. In fact, the differences observed in these 54
pairs are: $\Delta d\leq 1.0\AA$, $\Delta\theta\leq 0.3$, and $\Delta\tau\leq 0.6$.
We expect that when these H-forms are compared between different families those
leading in packing will have a larger chance to be shared than those dragged-in.

We next examine four domains from the other two families: d1id0a, d1i58a of
d.122.1.3, and d1s14a, d1h7sa2 of d.122.1.2. The situation within a single
family is similar to that in family d.122.1.4. There are 18 aligned pairs of similar
H-forms between d1id0a and d1i58a, and 15 between d1s14a and d1h7sa2. Taking domain
%d.122.1.3 81 H-forms, 19 aligned
%d.122.1.2 56 H-forms, 16 aligned
d1gkza2 as the template, which mainly consists of three long helices ($h_1$ to $h_3$) and one
sheet of five strands ($\beta_1$ to $\beta_5$, arranged as $h_1\beta_1h_2\beta_2\beta_3h_3
\beta_4\beta_5$), we can align all the other five domains against d1gkza2 with a large aligned
portion. The common region overlaps with $h_2\beta_2\beta_3h_3\beta_4\beta_5$. We find
that many pairs of similar H-forms are shared among the six domains. Three common H-forms
(of types HH, HE and HE) of the three families are shown in Table 4. %HH,HE,HE
%(The second fragment of H-form 1 of d1h7sa2 has a negative sequence score with that of
%d1gkza2, but has a positive sequence score with those of other domains. Since the BLOSUM
%similarity score is template-dependent a better way to measure similarity is to use
%weight matrices.)

%1gkza2: GROUP: 1as4.1\#A114 : 10 12  13,  69  72, HH  1  12 QEVID|KELVE;  1   3 SVVNA|EALSS
In fact, there are other H-forms shared at the superfamily level, e.g. 15 KIIEK - 57 YILPE
and 137 YAEYL - 142 GGSLQ of d1gkza2, which are respectively responsible for packing
$h_1h_2$ and $h_2\beta_4$ of d1gkza2.

\subsection{A global packing motif as a combination of packing units}
We have further examined SCOP40 family c.2.1.2 by taking domain d1e7wa as the template.
Roughly speaking, the structure of d1e7wa is relatively simple, consisting of a main
sheet of seven strands ($\beta_1$ to $\beta_7$) and six helical regions ($h_1$ to $h_6$)
between every two successive strands. There are four styles of H-forms ($h_1h_2$, $h_1h_6$,
$h_3h_4$, and $h_4h_5$) responsible for packing of helices, and nine styles related to
supersecondary structures. The remaining four styles of H-forms between helices and the sheet
are $\beta_2h_3$, $\beta_4h_6$, $\beta_5h_6$ and $h_5\beta_7$. Packing units are identified
as H-forms shared by many members of a family or even across families. Some representative
packing units identified for d1e7wa are\\
{\tt\phantom{44}$h_1h_2$ 15 LGRSI - 42 NALSA; $h_1h_6$ 17 RSIAE - 239 DVVIF; $h_3h_4$ 85 LVAAC - 135 IKAFA; \\
\phantom{44}$h_4h_5$ 121 ADLFG - 171 YTIYT; $\beta_1h_1$ 6 ALVTG - 18 SIAEG; $h_2\beta_2$ 18 SIAEG - 28 YAVCL;\\
\phantom{44}$\beta_2h_3$ 29 AVCLH - 87 AACYT; $\beta_4h_6$ 98 VLVNN - 239 DVVIF; $\beta_5h_6$ 155 SIINM -242 IFLCS.}\\

We have also examined domain d1u0sy of SCOP c.23.1.1. Its structure consists of a single
sheet of five strands and five intervened helices. There are three styles of H-forms
($h_1h_5$, $h_2h_3$, and $h_3h_4$) responsible for packing of helices, nine styles related to
supersecondary structures, and five other styles between helices and the sheet. We have
identified packing units for d1u0sy by inspecting also some members of c.23.1.1 and
c.23.1.2.\ or c.23.1.3. For example, the representative packing units for helices are\\
{\tt\phantom{44}$h_1h_5$ 20 DIITK - 110 RVVEA; $h_2h_3$ 34 TNGRE - 63 IDAIK;
$h_3h_4$ 61 NGIDA - 91 IEAIK.}\\

The size of d1u0sy is less than half of that of d1e7wa, but a large proportion of d1u0sy
can be aligned to d1e7wa by DALI at RMSD of 2.7\AA . Thus, we expect that they would
share some packing units. Four such units are listed in Table 5. The first two H-forms are
supersecondary structures, which might be formed under a mechanism more or less different
from that for SSEs distant in sequence, and then exhibit also some conservation in their
connecting loops.

After packing units of domains (or families) have been identified, their
H-forms as consisting elements may be merged in groups according to their
similarity in sequence and local geometry. The packing units just found are such
examples. We may then describe a global packing motif as a combination of
packing unit groups.

\subsection{Structure alignment based on similar H-forms}
We select three pairs of similar H-forms from domains d1gkza2 and d1id0a:\\
{\tt\phantom{44}1, 55 LDYIL - 132 PTSRA $\sim$ 46 FVEVM - 114 AVARE, (0 4; 0.6  0.4 -0.2 0.0);\\
\phantom{44}2, 25 RRLCE - 62 LLKNA $\sim$ 16 SALNK - 53 VLDNA, (0 14; 1.3,  0.3  0.1  0.1); \\
\phantom{44}3, 65 NAMRA - 82 PDVVI $\sim$ 114 AVARE- 125 GKIVA; (1 3; -0.5,  0.4 -0.2  0.1).}\\
The first pair is consistent with the DALI alignment of the two domains while the second has
a shift in SS1. (Residue L25 of d1gkza2 should align to V19 of d1id0a instead of L16.) The third
has no correspondence to the DALI alignment. As one can expect, indeed, at the first case
the alignment based on the pair of H-forms agrees with the DALI alignment regardless of
whether a zoom-in is performed or not.

The pairs of similar H-forms with a shift at one helix in comparison with the global
alignment are often seen. The second case is an example. The alignment based on the pair is\\
{\tt\phantom{44}d1gkza2: \underline{RRLCE}-YILPE\underline{LLKNA}MR-RISDR-GTDVY,\\
\phantom{44}d1id0a:\phantom{30} \underline{SALNK}-EVMGN\underline{VLDNA}CK-VVEDD-GARME},\\
where the deviation cutoff is 3.0\AA, the minimal width of aligned segments is 4, and
the H-forms taken for the initial alignment are underlined. If a
zoom-in is conducted the global alignment can be still recovered. However, in the third case
the final alignment is almost only the initial H-forms themselves no matter with or without
a zoom-in. The alignment without a zoom-in is\\
{\tt\phantom{44}d1gkza2:  \underline{NAMRA}T-V\underline{PDVVI}T,\\
\phantom{44}d1id0a:\phantom{30} \underline{AVARE}I-E\underline{GKIVA}G}.

We have also take take the following pair from Table 5\\
{\tt\phantom{44}1e7wa: 17 RSIAE 239 DVVIF $\sim$ 1u0sy: 16 MMLKD 109 SRVVE}\\
as initial alignment to align the two domains. The final alignment is\\
{\tt\phantom{44}1e7wa: PVALVTG-G\underline{RSIAE}GLHAEGYAV-CDVLVN-IINMVD-VNGVG-\underline{DVVIF}LCS\\
    \phantom{44}1u0sy: KRVLIVD-R\underline{MMLKD}IITKAGYEV-PDIVTM-IIVCSA-KDFIV-\underline{SRVVE}ALN}.\\
Indeed, many fragments including all the SSEs list in Table 5 are consistent with the H-form in
the structure alignment.

%The alignment between 1e7wa and 1dbwa has been conducted based on the following H-forms\\
%{\tt\phantom{44}1e7wa: 155 SIINM 242 IFLCS $\sim$ 1dbwa: 78 PSIVI 113 IEAIE}\\
%which are H-forms 4 in Table 5. The alignment \\
%{\tt\phantom{44}1e7wa: PVALVTG-G\underline{RSIAE}GLHAEGYAV-CDVLVN-IINMVD-VNGVG-\underline{DVVIF}LCS?\\
%    \phantom{443}1dbw: KRVLIVD-R\underline{MMLKD}IITKAGYEV-PDIVTM-IIVCSA-KDFIV-\underline{SRVVE}ALN?}\\
%shows the packing of the two helices ($h_1h_5$).

\subsection{A comparison with LDs}
A kind of multifragment structure motifs is the so-called local
descriptors (LDs).\cite{ld09}
An LD has a center residue and several fragments of width at least five which
closely surround the center in 3D space but not necessarily near each other
along the protein sequence. With a LD as a seed, many structurally similar LDs are
organized as a group, and then a library may be built from such groups.
A good library of LDs should have many possible applications, including protein
structure analysis, classification, alignment, identification of structure
domains, and structure and function prediction.

With ``GROUP: 1lara1\#1574: 7 ''taken as an example, its seed LD is from domain
d1lara1, and the seed center is residue I1574 of the domain.
This group have seven members, each of which consists of four
fragments (of widths 10, 9, 5 and 5, respectively). Except the third fragment,
which is on a loop, the other three are on three different helices; the center
of each LD sits on the fourth fragment. If two closely related
structures can be aligned against each other at totally 29 sites there should be
a good chance for the alignment to coincide with the pairwise structure alignment.
The seven members come from six protein domains, namely 1fpza\_\#189, 1jlna\_\#532,
2shpa1\#514, 1g4us2\#527,  1lara2\#1862, 1lara2\#1865 and 1lara1\#1574. 
In fact, two members 1lara2\#1862 and 1lara2\#1865 come from the same domain 1lara2, 
Three fragments from 1lara2\#1862, 1lara2\#1865 are the same, 
while that the center sits shift three residues generate that
 two similar local descriptors in the same group. 
Another redundancy is that even the same central residue may give rise to two different local descriptors,
 though both are very similar. The member 1jlna\_\#532 above have four fragments,
another local descriptor have three fragments, 
which also called 1jlna\_\#532 in ``GROUP: 1qgra\_\#821 : 1917'' 

Taking 1lara1 as the template,
we align the other five domains against 1lara1 with DALI.\cite{dali}
Indeed, four LDs agree with the DALI alignments of domain structures, but two
LDs have the fragments where the center sits ({\tt DQYQL} of 1lara2 and
{\tt SQFVQ} of 1g4us2) shifted roughly by a helical turn. The shifts are small
in 3D space, but cause a large drop in sequence similarity.

We list here the first and fourth fragments of 1lara1 and 1g4us2 of this group: \\
{\phantom{44} \tt 1lara1  RTGCFIVIDA - VFIHE $\sim$ 1g4us2  RTGTMAAALV - SQFVQ.}\\
A pair of H-forms between the two helices related to these two fragments are\\
{\phantom{44} \tt 1lara1   TGCFI - EDQYV $\sim$ 1g4us2 TGTMA - ASQFV \quad
(9, 11; 0.1, 0.1 $-0.3$  0.3).}\\
The BLOSUM62 similarity score between {\tt VFIHE} and {\tt SQFVQ} of LDs is $-6$,
while that between {\tt EDQYV} and {\tt ASQFV} of H-forms is 11. We expect that
the sequence signal for the latter would be much stronger. As shown, the
differences in distance $d$ and in three angles are all small. The RMSD tolerance
used for grouping LDs seems not sensitive enough. Moreover, since LDs are
collected based on a central residue it is often seen that many LDs belong to
the same structure alignment.

\section{Conclusions}

Our argument about the existence of packing units for the structure is in logic rather
than in causality. From the examined examples we have seen that there are many H-forms
similar in the local geometry between two structures in a same family. However, the
pairs of H-forms with also sequence similarity are much fewer, and those consistent
with the structure alignment are even fewer. The pairs of similar H-forms shared by
several family members or even among families should play a fundamental role in
packing although our analysis does not involve the physical interaction in H-forms.
Such H-forms are conservative in both structure and sequence, serve as packing units,
and carry strong sequence signals. After extracting such packing units from different
families, we may further organize them into a database library. A structure database
will map to a network of packing unit groups linked by the SSEs shared among packing
units. This is under study.

From the viewpoint of packing units, contacts are a mixture of those leading in
packing and those dragged-in. The latter carry much weaker sequence signal than the
former, so impede a reliable prediction. The former separated from the latter
would provide trusty 3D constraints for the structure prediction and structure
annotation. So far, we have not discussed the formation of a sheet from strands. A
primitive observation indicates that there are key contacts which are more
conservative than other contacts of hydrogen bonding pairs among strands of a sheet.

We have described a way to do structure alignment based on a pair of highly similar
H-forms. The number of pairs of highly similar H-forms between two protein structures
usually is not very large. Thus, we may develop a tool for structure alignment by
taking a pair of highly similar H-forms as a trial to initiate the alignment.

The packing units discussed above are mainly for domains, but they are valid for
analyzing packing across domains and even at the interface of a protein
complex.\cite{dock}
The packing units across two structures of a protein complex help us understand
the docking. The structure alignment based on a similar pair of such H-forms can
be developed for that purpose.

%linchpin 2chf D11,D56,K108?

\begin{quotation}
{This work is supported by the National Natural Science Foundation of
China and the National Basic Research Program of China (2007CB814800).
}
\end{quotation}

% see \cite{kn:gnus} for the detail -> see [67] see for the detail

\bigskip
\bigskip
\bigskip
\bigskip
\bigskip
\bigskip
\bigskip
\bigskip
\bigskip
\bigskip
\bigskip
\bigskip
\bigskip
\bigskip
\bigskip
\bigskip
\bigskip
\bigskip
\bigskip
\bigskip
\bigskip
\bigskip
\bigskip
\bigskip
\bigskip

\begin{center}\parbox{15cm}{Table 1. The H-forms of CI2. Pos1: the position of the
center of SS1; Pos2: that of SS2; Type: types of the two SSEs;
$d$: distance between the $C_\alpha$ atoms at Pos1 and Pos2; $\theta_a$, $\theta_b$:
the angles between the joint direction from Pos1 to Pos2 and the two SSE axes;
$\tau_{ab}$: dihedral angle made by the three directions. Distance is in \AA units
and angle in radian.}

\smallskip
%{\setlength{\tabcolsep}{1mm}\tt %\small\tt%\begin{tabular}
\begin{tabular}{r|ccc|cccc}
\hline
&Pos1& Pos2& Type& $d$& $\theta_a$& $\tau_{ab}$& $\theta_b$\\
\hline
1& \phantom{5}8&  16& HH& 6.4&  1.4&\phantom{5}$ 0.2$&  1.6\\
2& 13&  31& HE& 8.0&  1.6& \phantom{5} $0.1$&  1.3\\
3& 13&  49& HE& 7.6&  1.4& $-0.2$&  1.9\\
4& 13&  51& HE& 7.9&  2.2& \phantom{5}$0.0$&  1.3\\
5& 16&  49& HE& 8.4&  1.8& $-0.3$&  1.5\\
6& 17&  29& HE& 7.3&  1.4& $-0.5$&  1.6\\
7& 20&  47& HE& 8.1&  1.7& $-0.3$&  1.6\\
\hline\end{tabular}\end{center}

\begin{center}\parbox{15cm}{Table 2. The H-forms of 1vbw corresponding to those of CI2.
Sim1: the BLOSUM62 similarity score between the SS1s of the corresponding H-forms;
Sim2: that between the SS2s; the last four columns are the differences of
distance and angles of CI2 from those of corresponding ones in 1vbw (see the caption
of Table 1).}

\smallskip
\begin{tabular}{r|rrrr|ccrr}
\hline
&Pos1& Pos2& Sim1& Sim2& $\Delta d$& $\Delta\theta_a$& $\Delta \tau_{ab}$&
$\Delta \theta_b$\\
\hline
1& 11&  19& 23&   6& 0.4&  0.0& $ 0.0$& $ 0.1$\\
2& 16&  34& $-$4&   6& 1.6&  0.2& $-0.3$& $ 0.1$\\
3& 16&  52& $-$4&  15& 1.0&  0.2& $0.2$& $-0.1$\\
4& 16&  54& $-$4&   6& 1.4&  0.0& $-0.2$& $ 0.0$\\
5& 19&  52&  6&  15& 0.8&  0.0& $ 0.1$& $-0.1$\\
6& 20&  32& 11&   6& 1.9&  0.3& $-0.3$& $ 0.0$\\
7& 23&  50&  5&  21& 0.1&  0.0& $ 0.0$& $-0.1$\\
\hline\end{tabular}\end{center}

\begin{center}\parbox{15cm}{Table 3. The H-forms of 1mit corresponding to those of CI2
(see the caption of Table 2).}

\smallskip
\begin{tabular}{r|rrrr|rrrr}
\hline
&Pos1& Pos2& Sim1& Sim2& $\Delta d$& $\Delta\theta_a$& $\Delta \tau_{ab}$&
$\Delta \theta_b$\\
\hline
1& \multicolumn{4}{c|}{$-$}&  \multicolumn{4}{c}{$-$}\\
2& 17& 35& $-7$& $ 8$& $ 2.0$& $ 0.2$& $ 0.2$& $-0.2$\\
3& 17& 53& $-7$& $16$& $ 0.8$& $ 0.2$& $0.6$& $ 0.3$\\
4& 17& 55& $-7$& $13$& $ 0.7$& $-0.1$& $0.3$& $ 0.5$\\
5& 20& 53& $ 6$& $16$& $ 0.9$& $ 0.2$& $0.7$& $-0.1$\\
6& 21& 33& $ 9$& $ 6$& $ 2.6$& $ 0.4$& $0.4$& $0.0$\\
7& 24& 51& $ 4$& $17$& $ 1.5$& $ 0.0$& $ 0.1$& $ 0.2$\\
\hline\end{tabular}\end{center}

\begin{center}\parbox{12cm}{Table 4. The H-forms shared among domains of SCOP-40
d.122.1.}

\smallskip{\tt
\begin{tabular}{l|rr|rr|rr}
\hline
&\multicolumn{2}{c|}{H-form 1}&\multicolumn{2}{c|}{H-form 2}&\multicolumn{2}{c}{H-form 3}\\
\hline
{d1gkza2}&60 PELLK& 132 PTSRA&64 KNAMR& 82 PDVVI& 64 KNAMR& 153 IGTDV\\
{d1y8oa2}&71 FELFK& 139 PISRL&75 KNSMR& 93 PAVKT &75 KNSMR& 160 VGTDA\\
{d1id0a}&51 GNVLD& 114 AVARE&55 DNACK&  63 EFVEI& 55 DNACK& 135 GGARM\\
{d1i58a}&54 LHLLR& 158 DVVKN&58 RNAID& 80 GTLIL& 58 RNAID& 179 KGTKV\\ %58 80 new
{d1s14a}&12 QEVID& 69 SVVNA&16 DNSVD&  28 KRVDV& 16 DNSVD& 113 TGTSV\\
{d1h7sa2}&\multicolumn{2}{c|}{$-$}&17 ENSLD&  25 TNIDL& 17 ENSLD& 116 RGTTV\\ %13 KELVE& 72 KELVE
\hline\end{tabular}}\end{center}

\begin{center}\parbox{12cm}{Table 5. The H-forms shared between domains of SCOP-40 c.2.1 and c.23.1.}

\smallskip{\setlength{\tabcolsep}{1mm}\small\tt
\begin{tabular}{l|l||rr|rrrr|rr|rrrr}
\hline
\multicolumn{2}{c||}{SCOP domain}&\multicolumn{2}{c|}{H-form 1 (EH)}&$d$&$\theta_a$&$\tau$&$\theta_b$&
\multicolumn{2}{c|}{H-form 2 (HE)}&$d$&$\theta_a$&$\tau$&$\theta_b$\\
\hline
        & 1e7wa &  6 ALVTG&  18 SIAEG& 6.5& 1.5&  0.5& 1.2&   18 SIAEG&  28 YAVCL& 6.6& 1.0&$ 0.5$& 1.7\\
        & 1e6ua &  6 VFIAG&  18 AIRRQ& 7.2& 1.7& $-$0.2& 1.1&   18 AIRRQ&  29 VELVL& 8.0& 1.1&$ 0.6$& 1.7\\
c.2.1.2 & 1bdba &  9 VLITG&  21 ALVDR& 7.6& 1.4&  0.6& 1.2&   21 ALVDR&  31 AKVAV& 6.6& 1.0&$ 0.6$& 1.7\\
        & 1fjha &  5 IVISG&  17 ATRKV& 7.3& 1.4&  0.8& 1.2&   17 ATRKV&  27 HQIVG& 7.0& 1.1&$ 0.5$& 1.6\\
        & 1hxha & 10 ALVTG&  22 EVVKL& 6.9& 1.6&  0.2& 1.1&   22 EVVKL&  32 AKVAF& 6.5& 1.0&$ 1.0$& 1.9\\
\hline
c.2.1.5 & 1y7ta &  8 VAVTG&  20 SLLFR& 8.3& 1.3&  1.0& 1.2&   20 SLLFR&  37 VILQL& 8.2& 1.1&$ 0.3$& 1.6\\
\hline
        & 1dbwa &  7 VHIVD&  19 SLAFM& 6.9& 1.6&  0.1& 1.2&   19 SLAFM&  29 FAVKM& 6.5& 0.9&$ 1.6$& 1.9\\
c.23.1.1& 1u0sy &  5 VLIVD&  17 MLKDI& 8.0& 1.8& $-$0.1& 1.1&   17 MLKDI&  27 YEVAG& 6.6& 1.2&$ 0.1$& 1.1\\
        & 1s8na &  7 VLIAE&  19 DLAEM& 7.7& 1.5&  0.1& 1.3&  \multicolumn{6}{c}{$-$}  \\
        & 1kgsa2&  5 VLVVE&  17 LITEA& 6.8& 1.7&  0.0& 1.2&   17 LITEA&  27 FTVDV& 6.3& 1.0&$ 1.4$& 2.1\\
\hline
c.23.1.2& 1dcfa & 11 VLVMD&  23 VTKGL& 7.1& 1.9& $-$0.4& 1.2&   23 VTKGL&  33 CEVTT& 6.1& 1.1&$ 0.8$& 1.9\\
\hline\hline
\multicolumn{2}{c||}{SCOP domain}&\multicolumn{2}{c|}{H-form 3 (HH)}&$d$&$\theta_a$&$\tau$&$\theta_b$&
\multicolumn{2}{c|}{H-form 4 (EH)}&$d$&$\theta_a$&$\tau$&$\theta_b$\\
        \hline
        & 1e7wa & 17 RSIAE& 239 DVVIF& 7.0& 1.4&  2.4& 1.1&  155 SIINM& 242 IFLCS& 7.9& 2.0&$-0.3$& 1.3\\
        & 1e6ua & 17 SAIRR& 219 AASIH& 7.8& 1.7&  2.0& 1.1&  101 KLLFL& 222 IHVME& 8.2& 1.5&$-0.9$& 1.5\\
c.2.1.2 & 1bdba & 20 RALVD& 222 GAYVF& 7.9& 1.4&  2.2& 1.0&  137 NVIFT& 225 VFFAT& 7.1& 1.9&$-0.7$& 1.3\\
        & 1fjha & 16 AATRK& 204 SVIAF& 7.2& 1.5&  2.3& 1.0&  109 AAVVI& 207 AFLMS& 7.2& 2.0&$-0.7$& 1.4\\
        & 1hxha & 21 LEVVK& 225 QLVLF& 6.9& 1.5&  2.4& 1.1&  133 SIINM& 228 LFLAS& 7.1& 2.1&$-0.2$& 1.4\\
\hline
c.2.1.5 & 1y7ta & 19 YSLLF& 244 NAAIE& 8.0& 1.6&  2.4& 1.1&  126 KVLVV& 247 IEHIR& 8.0& 1.5&$-1.4$& 1.6\\
\hline
        & 1dbwa & \multicolumn{6}{c|}{$-$}&                   78 PSIVI& 113 IEAIE& 7.1& 1.6&$-1.0$& 1.8\\
c.23.1.1& 1u0sy & 16 MMLKD& 109 SRVVE& 8.4& 1.0&  1.8& 1.3&   77 KIIVC& 112 VEALN& 7.4& 1.5&$-0.8$& 1.9\\
        & 1krwa & \multicolumn{6}{c|}{$-$}&                   78 PVIIM& 113 VALVE& 7.9& 1.4&$-0.5$& 1.5\\
        & 1kgsa2& 16 DLITE& 108 RELIA& 8.3& 1.3&  2.0& 1.3&   76 PVLML& 111 IARVR& 7.7& 1.5&$-0.8$& 1.7\\
\hline
c.23.1.3& 1qo0d & 26 DALVL& 113 HRVLP& 7.8& 1.0&  1.9& 1.2&   81 TLVAL& 116 LPVLV& 8.0& 1.8&$ 0.1$& 1.9\\
\hline\end{tabular}}\end{center}


\begin{thebibliography}{99}

\bibitem{brutlag04} Shapiro J,  Brutlag D. FoldMiner: Structural motif
discovery using an improved superposition algorithm. Protein Sci 2004;13:278-294.
\bibitem{sander06} Sander O,  Sommer I,  Lengauer T. Local protein structure prediction using
discriminative models. BMC Bioinformatics 2006;7:14 p.1-13.
\bibitem{baker01} Bonneau R,   Baker D. Ab Initio Protein Structure Prediction: Progress
and Prospects. Rev Biophys Biomol Struct 2001;30:173-89.
\bibitem{smotif10} Fernandez-Fuentes N,  Dybas JM,  Fiser A. Structural characteristics of novel
protein folds. PLoS Comput Biol 2010;6:e1000750 p.1-11. (Smotifs)
\bibitem{tops} Michalopoulos I,  Torrance GM,  Gilbert DR,  Westhead DR.
TOPS: an enhanced database of protein structural topology.
Nucleic Acids Res 2004;32:D251-D254.
\bibitem{prosmos} Shi SY,  Chitturi B,  Grishin NV. ProSMoS server: a pattern-based search
using interaction matrix representation of protein structures. Nucleic Acids Res 2009;37:W526-531.
\bibitem{baker99}Simons KT,  Ruczinski I,  Kooperberg C,  Fox BA,  Bystroff C,   Baker D.
Improved Recognition of Native-like Protein Structures using A Combination
of Sequence-dependent and Sequence-independent Features of
Proteins. Proteins 1999;34:82-95.
\bibitem{ld09} Hvidsten TR,  Kryshtafovych A,  Fidelis K. Local descriptors of protein structure:
A systematic analysis of the sequence-structure relationship in proteins using
short- and long-range interactions. Proteins 2009;75:870-884.
\bibitem{nature}Shakhnovich EI, Abkevich VA, Ptitsyn O. Conserved residues and
the mechanism of protein folding. Nature 1996;379:96-98.
\bibitem{broglia} Camilloni C,  Sutto L,  Provasi D,  Tiana  G,   Broglia RA. Early events
in protein folding: Is there something more than hydrophobic burst? Protein
Sci 2008;17:1424-1433.
\bibitem{dssp}Kabsch W, Sander C.  Dictionary of protein secondary structure: Pattern
recognition of hydrogen-boned and geometrical features. Biopolymers 1983;22:2577-2637.
\bibitem{contact}  da Silveira CH,  Pires DEV,  Minardi RC, {\it et al.}
%C Ribeiro, CJM. Veloso, JCD. Lopes, W Meira Jr., G Neshich, CHI. Ramos, R Habesch, MM. Santoro
Protein cutoff scanning: A comparative analysis of cutoff dependent and cutoff free
methods for prospecting contacts in proteins. Proteins 2009;74:727-743.
\bibitem{clepaps} Wang  S,  Zheng WM. CLEPAPS: Fast pair alignment of protein structures based
on conformational letters. J Bioinfor Comput Biol 2008;6:347-366.
%\potential\cutoff-voro.pdf
%\bibitem{pdb}
\bibitem{pfam} Sonnhammer ELL, Eddy SR,   Durbin R. Pfam: A comprehensive database of protein
domain families based on seed alignments. Proteins 1997;28:405-420.
\bibitem{dali}Holm L, Sander C. Dictionary of Recurrent Domains in Protein
Structures. Proteins 1998;33:88-96.
\bibitem{dock}Gao M,  Skolnick J. iAlign: a method for the structural comparison of
protein-protein interfaces. Bioinformatics 2010;26:2259-65.


\end{thebibliography}
\end{document}